\documentclass[sigconf]{acmart}

\usepackage{bm}
\usepackage{subfig}
\usepackage{multirow}
\usepackage{diagbox}
\usepackage{enumitem}

\AtBeginDocument{%
  \providecommand\BibTeX{{%
    \normalfont B\kern-0.5em{\scshape i\kern-0.25em b}\kern-0.8em\TeX}}}

\copyrightyear{2022}
\acmYear{2022}
\setcopyright{rightsretained}
\acmConference[WSDM '22] {Proceedings of the Fifteenth ACM International Conference on Web Search and Data Mining}{February 21--25, 2022}{Tempe, AZ, USA.}
\acmBooktitle{Proceedings of the Fifteenth ACM International Conference on Web Search and Data Mining (WSDM '22), February 21--25, 2022, Tempe, AZ, USA}
\acmPrice{}
\acmISBN{978-1-4503-9132-0/22/02}
\acmDOI{10.1145/3488560.3498435}

\begin{document}

\title{CAN: Feature Co-Action for Click-Through Rate Prediction}
\author{Weijie Bian$\dagger$, Kailun Wu$\dagger$, Lejian Ren$\dagger$, Qi Pi$\dagger$, Yujing Zhang$\dagger$, Can Xiao$\dagger$}

\author{Xiang-Rong Sheng$\dagger$, Yong-Nan Zhu$\dagger$, Zhangming Chan$\dagger$, Na Mou$\dagger$, Xinchen Luo$\dagger$}
\author{Shiming Xiang$\ddagger$, Guorui Zhou$\dagger$, Xiaoqiang Zhu$\dagger$, Hongbo Deng$\dagger$}
\authornote{Hongbo Deng is the corresponding author (dhb167148@alibaba-inc.com).}

\affiliation{%
  \institution{$\dagger$Alibaba Group\qquad \qquad \qquad$\ddagger$Institute of Automation, Chinese Academy of Sciences}
  \country{China}
}
\email{{weijie.bwj,kailun.wukailun,lejian.rlj,dhb167148}@alibaba-inc.com}

\renewcommand{\shortauthors}{Weijie Bian, et al.}
\fancyhead{}


\begin{abstract}
  Feature interaction has been recognized as an important problem in machine learning, which is also very essential for click-through rate (CTR) prediction tasks. In recent years, Deep Neural Networks (DNNs) can automatically learn implicit nonlinear interactions from original sparse features, and therefore have been widely used in industrial CTR prediction tasks. However, the implicit feature interactions learned in DNNs cannot fully retain the complete representation capacity of the original and empirical feature interactions (e.g., cartesian product) without loss. For example, a simple attempt to learn the combination of feature A and feature B <A, B> as the explicit cartesian product representation of new features can outperform previous implicit feature interaction models including factorization machine (FM)-based models and their variations. This indicates there is still a big gap between explicit and implicit feature interaction models. However, to learn all the explicit feature interaction (cartesian product) representations requires a very large sample size along with $N$ times of original parameter space (where $N$ is quite large in most industrial applications). In this paper, we propose a Co-Action Network (CAN) to approximate the explicit pairwise feature interactions without introducing too many additional parameters. More specifically, giving feature A and its associated feature B, their feature interaction is modeled by learning two sets of parameters: 1) the embedding of feature A, and 2) a Multi-Layer Perceptron (MLP) to represent feature B. The approximated feature interaction can be obtained by passing the embedding of feature A through the MLP network of feature B. We refer to such pairwise feature interaction as feature co-action, and such a Co-Action Network unit can provide a very powerful capacity to fitting complex feature interactions. In addition, FM can be viewed as a special case of the CAN unit when the MLP is a single layer with only one output. Experimental results on public and industrial datasets show that CAN outperforms state-of-the-art CTR models and the cartesian product method. Moreover, CAN has been deployed in the display advertisement system in Alibaba, obtaining 12\% improvement on CTR and 8\% on Revenue Per Mille (RPM), which is a great improvement to the business. The code for experiments in this paper is open-sourced\footnote{https://github.com/CAN-Paper/Co-Action-Network}.
\end{abstract}



\maketitle

\section{Introduction}\label{sec:intro}

With the growing complexity of machine learning models, especially those in recommendation systems, how to deal with abundant input features effectively and efficiently becomes a crucial problem. For online recommendation system in an industrial setting, models are often trained on billion-scale sparse features with one-hot encoding~\cite{cheng2016wide,zhou2018din,DBLP:conf/cikm/ShengZZDDLYLZDZ21}. Each feature can also be seen as a unique ID, which is often mapped to a low-dimensional embedding before being fed into the model. A simple way to deal with the large-scale input is to consider each feature independently. Under this strategy, a DNN can be directly trained to estimate CTR based on the combination (e.g., concatenation) of features in which the feature interaction relies on the implicit modeling of fully connected layers. 

\begin{figure}[t]
	\centering
	\includegraphics[width=0.4\linewidth]{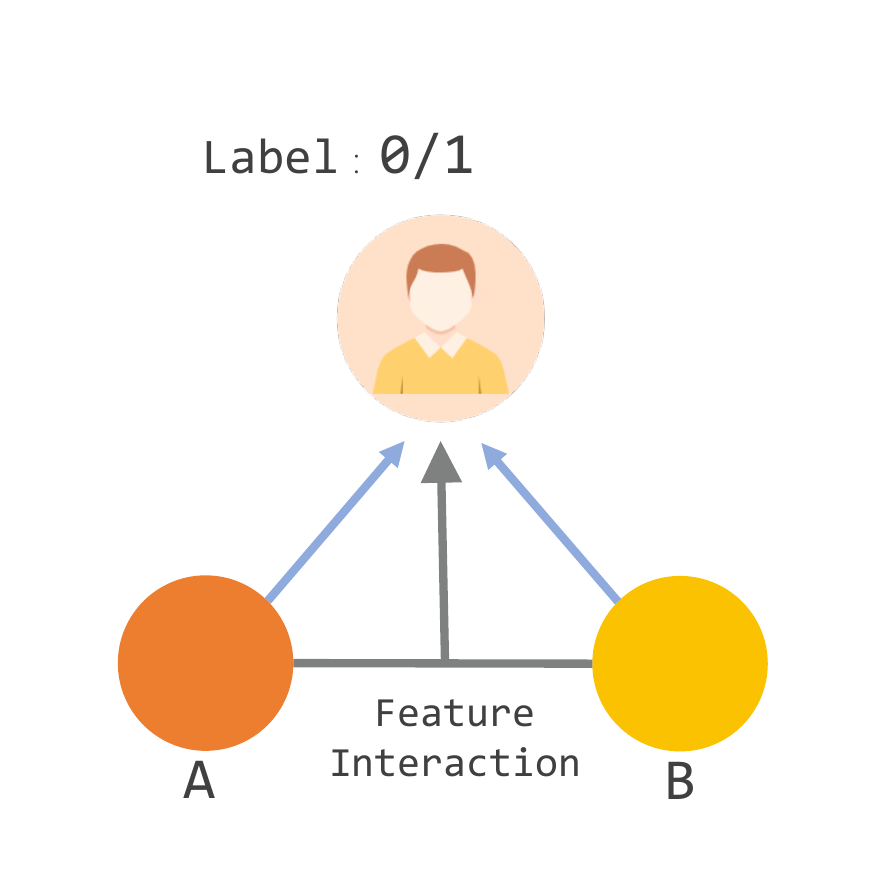}
	\vspace{-2mm}
	\caption{Illustration of feature interaction. Blue lines indicate the effects of two features A and B for the final label are modeled separately while feature interaction in the grey line bridges them together.}
	\label{fig:coaction}
\end{figure}

However, features like ``candidate item'' and ``user click history'' in the recommendation system are highly relevant~\cite{zhou2018din,zhou2019dien}, i.e., there exist co-occurrence information. Another typical example is the story of ``Beer and Nappies''. This kind of feature interaction is beneficial to better estimate CTR. As illustrated in Fig.\ref{fig:coaction}, if no feature interaction is considered, the effects of features A and B toward the final label (click or not in CTR prediction) are modeled independently, as shown by the blue lines. Feature interaction shown in the grey line explicitly bridges the correlations of feature pair (A, B) to the target label and brings more useful information for the learning objective. We will discuss it more deeply in Sec. \ref{sec:revisit}.

To model feature interaction, the most simple way is to use cartesian product. Given two features A and B, once features A and B are selected, the co-occurrence <A, B> is treated as a new feature and fed into the model. As the co-occurrence information is provided directly in the form of additional input, the training process gets easier. Although the cartesian product is simple and effective, it has some serious defects, such as huge parameter volume and feature space, poor generalization ability.

Different from cartesian product that introduces additional input information, several research efforts have been devoted to model feature interaction through careful combinations of input features. The typical examples are factorization based methods~\cite{rendle2010factorization,qu2016product,WangFFW2017DCN,lian2018xdeepfm,xiao2017attentional,song2019autoint} which emphasize both low- and/or high-order feature interactions by directly combining feature embeddings in latent vector space with various operators. As these methods consider feature interaction from model aspect and the operators are well-designed, they are usually more deployment-friendly compared with cartesian product. However, they still have some downsides that are nonnegligible. Regrettably, the shallow structures of some factorization based methods \cite{juan2016field,WangFFW2017DCN} limit their representative power. More critically, the embeddings generated by their operators take the responsibility of representation learning and feature interaction modeling simultaneously which may hinder the training process. This kind of combination reduces the memory ability of feature interactions, thus degrading the model capacity.

\author{Weijie Bian$\dagger$, Kailun Wu$\dagger$, Lejian Ren$\dagger$, Qi Pi$\dagger$, Yujing Zhang$\dagger$, Can Xiao$\dagger$}

\author{Xiang-Rong Sheng$\dagger$, Yong-Nan Zhu$\dagger$, Zhangming Chan$\dagger$, Na Mou$\dagger$, Xinchen Luo$\dagger$}
\author{Shiming Xiang$\ddagger$, Guorui Zhou$\dagger$, Xiaoqiang Zhu$\dagger$, Hongbo Deng$\dagger$}
\authornote{Hongbo Deng is the corresponding author (dhb167148@alibaba-inc.com).}

\affiliation{%
  \institution{$\dagger$Alibaba Group\qquad \qquad \qquad$\ddagger$Institute of Automation, Chinese Academy of Sciences}
  \city{Beijing}
  \country{China}
}
\email{{weijie.bwj,kailun.wukailun,lejian.rlj,dhb167148}@alibaba-inc.com}
To solve these issues, we propose \textbf{C}o-\textbf{A}ction \textbf{N}etwork (\textbf{CAN}) that can capture feature interaction and utilize the mutual and common information of different feature pairs effectively. Specifically, for each feature pair, the embeddings of one side (induction feature) are used to construct an MLP applied to the other side (feed feature) in the co-action unit. This paradigm of feature interaction modeling reduces the additional parameter from $O(N^2 \times D)$ to $O(N\times (D' + D))$ ($N$ is the number of features, $D$ and $D'$ are the dimensions of parameter used in co-action unit, $D < D' \ll N$) compared with cartesian product. Further, CAN provides more powerful fitting ability and better non-linearity using MLP in the co-action unit compared with traditional factorization based methods. Besides, CAN differentiates the parameter space for representation learning and feature interaction modeling to avoid mutual interference in training. Multi-order enhancement and multi-level independence are further utilized to enrich the expression ability in CAN. 

The main contributions of this work are summarized as follows:
\begin{itemize}[leftmargin=*]
\item We investigate the effectiveness of cartesian product and point out the potential of implicit feature interaction modeling. Inspired by the independent coding of cartesian product, we design a new feature interaction paradigm that has comparable performance compared with cartesian product yet has much less resource consumption.
\item We propose Co-Action Network (CAN) to model feature interaction among raw features at the input stage. Each feature ID in CAN will be distributed to an individual micro-MLP to model the interaction with other features. In this way, CAN improves the expressive ability of modeling feature interaction under limited parameters. The individual micro-MLP has a more powerful expressive ability than the ordinary operators utilized in factorization based methods. 
\item We conduct extensive experiments on both public and industrial datasets. The consistent superiority validates the effectiveness of CAN in modeling feature interaction compared with other state-of-the-art competitors. The deployment of CAN brings 12\% CTR and 8\% RPM lift in Alibaba's display advertisement system. 
\end{itemize}

\section{Related Work}
Several research efforts have been devoted to model feature interactions in CTR prediction. These methods can be divided into several categories: aggregation, graph and factorization based methods. We give a brief introduction and discussion as follows.

Deep CTR prediction models generally follow an Embedding \& MLP paradigm. Large-scale sparse input features, or IDs, are first mapped to low dimensional embedding vectors and then aggregated into fixed-length vectors in a group-wise manner. The finally concatenated vector is fed as the input to a multi-layer perceptron (MLP). A series of research work focuses on learning how to aggregate features to obtain discriminative representation for CTR prediction. DIN and DIEN~\cite{zhou2018din,zhou2019dien} uses attention to activate historical behaviors locally w.r.t. the given target item and successfully captures the diversity characteristics of user interest. MIND~\cite{Li2019MIND} uses sufficient multi-vectors to capture complicated patterns lying in the user and items. Moreover, inspired by the success of self-attention architecture in the tasks of sequence learning~\cite{VaswaniSPUJGKP2017Transformer,chan2020selection}, the transformer is introduced in~\citet{FengLSWSZY19DSIN} for feature aggregation. MIMN~\cite{Pi2019MIMN} proposes a memory-based architecture to aggregate features and tackle the challenge of long-term user interest modeling. These aggregation based methods only take feature interaction as the weight of each user action to represent user interest. 

Graph based methods like Graph Neural Networks (GNNs)~\cite{Gori2005GNN,hu2019gsn} conduct feature propagation for each node, where the neighborhood information is aggregated. ~\cite{BrunaZSL13SpectralGraph,KipfW17GCN,VelickovicCCRLB18GAT} propose a spectral graph-based extension of convolutional networks with a self-attention mechanism to graphs for feature propagation. There are also some works~\cite{SunHYYW2011PathSim,ShiHZY2019HINEmbedding,ZhaoYLSL2017HINFusion} that exploit meta-path between different nodes for embedding learning. Although graph-based methods achieve great success on graph data, feature interaction is modeled only by one-dimensional weight indicating the strength of connectives, resulting in insufficient expression of the interaction.

Factorization Machines (FM)~\cite{rendle2010factorization} is a representative method in the age of shallow models. In FM, feature interaction is modeled as the inner product of latent vectors of features. However, FM uses the same latent vectors in different types of inter-field interactions. DeepFM~\cite{guo2017deepfm} imposes a factorization machine as ``wide'' module in Wide\&Deep~\cite{cheng2016wide} with no need of constructing cartesian product feature manually. In Product-based Neural Network (PNN)~\cite{qu2016product,qu2018product}, a product layer is introduced to capture feature interactions between inter-field categories. Deep \& Cross Network (DCN)~\cite{WangFFW2017DCN} applies feature crossing at each layer. Operation-aware Neural Networks (ONN)~\cite{yang2020operation} learns feature interaction through some different operations. Although the above approaches achieve some performance gain compared with plain DNN, the embedding of each ID takes the responsibility of representation learning and interaction modeling simultaneously and the mutual interferences between them might hurt the performance.


\section{Revisiting Feature Interaction for CTR Prediction} \label{sec:revisit}

\begin{figure}[t]
	\centering
	\includegraphics[width=1\linewidth]{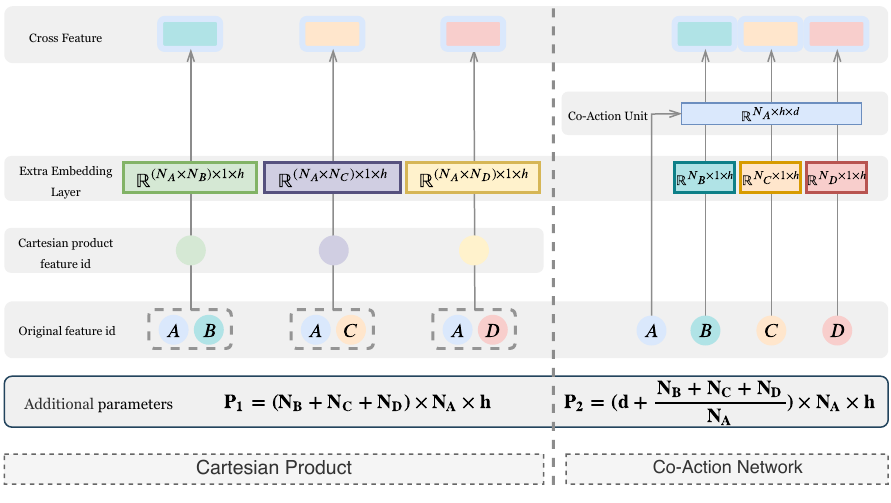}
	\caption{Illustration of the evolution from cartesian product to our feature co-action network, where $A, B$, $C$ and $D$ denote four kinds of the feature. 
$\text{N}_\text{A}, \text{N}_\text{B}$, $\text{N}_\text{C}$ and $\text{N}_\text{D}$ indicate the number of feature $A, B$, $C$ and $D$, respectively. $h$ is the dimension of the feature embedding and $d$ is the dimension of the output from co-action unit. In this figure, there we use feature $A$ to interact with other three features.}
	\label{fig:transportation}
\end{figure}

In the advertisement system, the predicted CTR $\hat{y}$ of a user $u$ clicking on an ad $m$ is calculated via: 
\begin{equation}
    \hat{y} = \mathrm{DNN} \big(E(u_1),\cdots,E(u_I),E(m_1),\cdots,E(m_J)\big)
\end{equation}
where $U = \{u_1,\dots,u_I\}$ is the set of user features including browsing and click history, user profile feature, etc., and $M = \{m_1,\dots,m_J\}$ is the set of item features. User and item features are usually unique IDs. $E(\cdot)\in \mathbb{R}^d$ means the embedding with size $d$ which maps the sparse IDs into learnable dense vectors as the inputs of DNNs. Besides these unary terms, previous works model feature interaction as binary terms:
\begin{equation}
	\begin{split}
		\hat{y} = &\mathrm{DNN} \big( E(u_1),\cdots,E(u_I),E(m_1),\cdots, E(m_J), \\
				 &\qquad \quad ~\{F(u_{i}, m_{j})\}_{ i \in [1, \cdots, I], j \in [1, \cdots, J]} \big)
	\end{split}
\end{equation}
where $F(u_{i}, m_{j}) \in \mathbb{R}^d$ represents the interaction between user feature $u_{i}$ and item feature $m_{j}$. 
The model can benefit from feature interaction due to the existence of feature co-occurrence, as shown in the example of ``Beer and Nappies'' in the previous section. Therefore, how to model the feature interaction effectively is crucial for improving performance.

After carefully revisiting previous methods, it can be found that they either take feature interaction as the weights or learn the correlation implicitly with other objectives simultaneously that may produce unsatisfactory results. The most direct way to learn feature interaction is to treat feature combinations as new features and learn an embedding vector for each feature combination directly, e.g., cartesian product. Cartesian product provides independent parameter space and thus has enough flexibility to learn co-action information to improve the ability of prediction.

However, there are some serious defects. The first one is the parameter explosion issue. 
The parameter space of cartesian product of two features with size $N$ will expand from $O(N \times D)$ to $O(N^2 \times D)$, where D is the dimension of embeddings, which will bring great burden to the online system. In addition, as cartesian product regards <A, B> and <A, C> as totally different features, there is no information sharing between combinations, which also limits the representation ability.  

\begin{figure*}[t]
	\centering
	\includegraphics[width=0.9\linewidth]{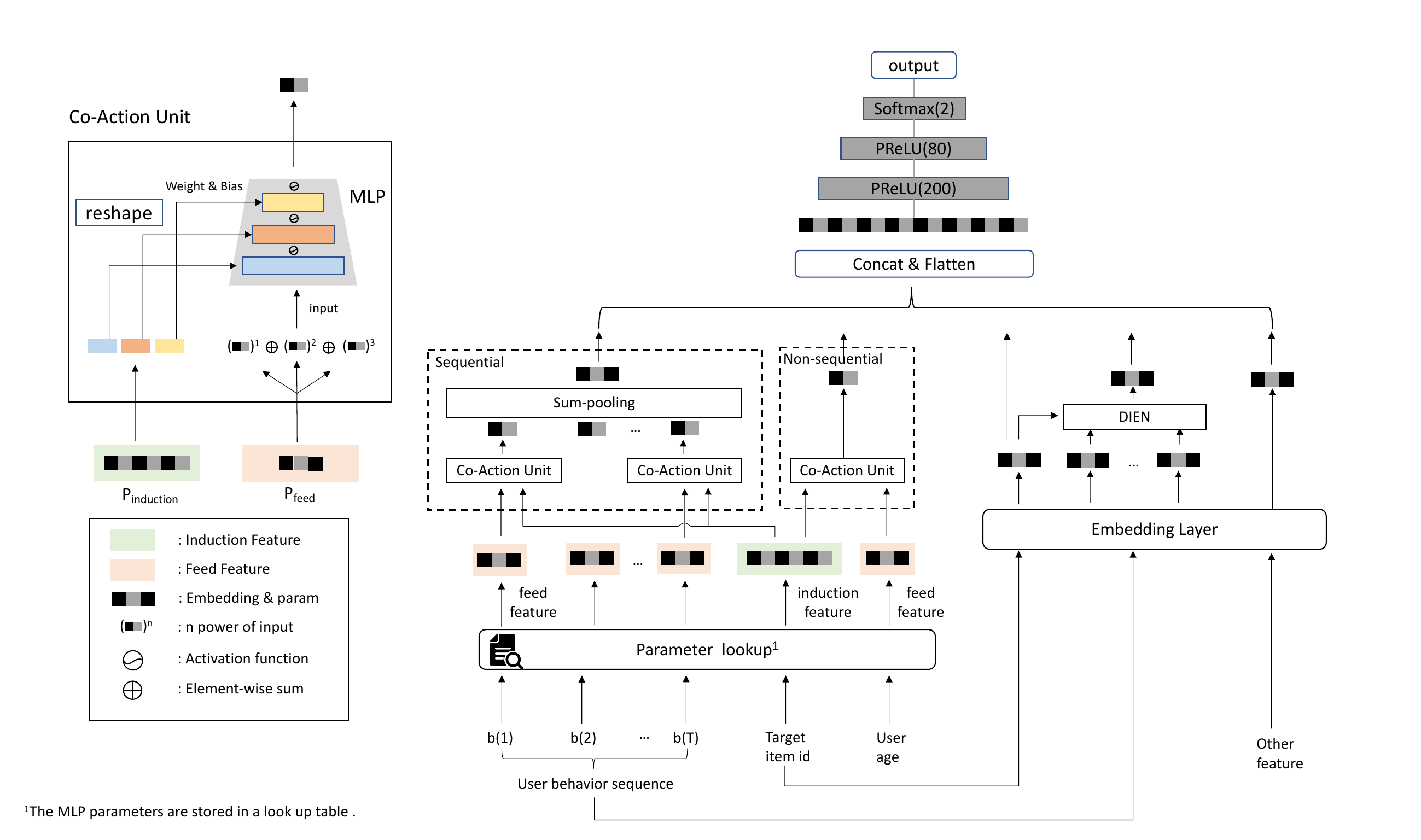}
	\caption{The overall framework of our Co-Action Network. Given target item and user features, the embedding layer encodes the sparse features into dense embeddings. Some selected features are divided into two sides $P_{induction}$ and $P_{feed}$ that are components of co-action unit. 
	$P_{induction}$ parameterizes the micro MLP and $P_{feed}$ serves as the input. The output of co-action unit, together with the common feature embeddings, are used to make final CTR prediction.}
	\label{fig:framework}
\end{figure*}

Taking into account both the advantages of cartesian product and the serving efficiency of calculation, we introduce a new way to model feature interaction. As illustrated in Fig.\ref{fig:transportation}(a), for each feature pair, its cartesian product produces a new feature and corresponding embedding. Since different feature pairs may share the same feature, there exists an implicit similarity between any two feature pairs, which is ignored by cartesian product.
If the implicit similarity can be effectively dealt with, the feature interactions among these pairs could be modeled more effectively and efficiently with a smaller parameter scale than cartesian product. In this paper, inspired by the independent coding of cartesian product, we firstly differentiate the parameters for embedding and feature interaction so that mutual interference is avoided. Considering that DNNs have the powerful fitting ability, we design a co-action unit that parameterizes the feature embeddings in the form of a micro-network. As different feature pairs can share the same micro-network, the similarity information is learned and stored naturally in this micro-network as illustrated in Fig.\ref{fig:transportation}(b). 


\section{Co-Action Network}\label{sec:method}

In this section, we propose Co-Action Network (CAN) to efficiently capture feature interaction which firstly introduces a pluggable module, co-action unit. The unit differentiates the parameters for embedding and feature interaction learning. Specifically, it is made up of two sides of information from raw features, i.e., induction and feed side. The induction side is used to construct a micro-MLP while the feed side provides the input for it. Moreover, to promote more non-linearity and deeply excavate feature interaction, multi-order enhancement and multi-level independence are introduced.

\subsection{Architecture Overview} \label{subsec:emb_as_mlp}
The whole architecture of CAN is shown in Fig.\ref{fig:framework}. The features $U$ and $M$ of a user and the target item are fed into CAN in two manners. In the first manner, they are encoded using embedding layer to dense vectors $\{E(u_1),\cdots,E(u_I)\}$ and $\{E(m_1),\cdots,E(m_J)\}$ and further concatenated as $e_{item}$ and $e_{user}$, respectively. In the second manner, we select a subset  $U_{feed}$ and $M_{induction}$ from $U$ and $M$ to model feature interactions $\{F(u_i, m_j)\}_{ u_i \in U_{feed}, m_j \in M_{induction}}$ using our proposed co-action unit. The detailed explanation and implementation of co-action unit will be elaborated in the next sub-section. The formulation of CAN is:

\begin{align}
\hat{y}=\mathrm{DNN}\big(e_{item},e_{user},\{F(u_i, m_j)\}_{ u_i \in U_{feed}, m_j \in M_{induction}}|\Theta)
\end{align}
where $\Theta$ denotes the parameters in the model and $\hat{y} \in [0, 1]$ is the predicted probability of the click behavior. The ground truth click information is denoted as $y \in \{0,1\}$. We finally minimize the cross-entropy loss function  between the prediction $\hat{y}$ and label $y$:

\begin{equation}
    \min_{\Theta}-y\mathrm{log}(\hat{y})-(1-y)\mathrm{log}(1-\hat{y})
\end{equation}

\subsection{Co-Action Unit} 
\label{subsec:co-action_unit}

Generally speaking, the co-action unit is an independent MLP for each feature pair, namely micro-MLP with the weight, bias and input of MLP provided by the feature pair.
For a specific user feature ID $u_{o'} \in U_{feed}$, we use parameter lookup to obtain learnable parameters $P_{induction} \in \mathbb{R}^{D'}$ while item feature ID $m_o \in M_{induction}$ for $P_{feed} \in \mathbb{R}^{D}$ ($D < D'$). Next, $P_{induction}$ is reshaped and split into the weight matrix and bias vector for the micro-MLP. This process can be formularized as:

\begin{align}
    &{\Big\Arrowvert_{i=0}^{L-1}} (w_i \| b_i) = P_{induction} \\
    &\sum_{i=0}^{L-1} (|w_i| + |b_i|) = | P_{induction} | = D'
\end{align}
where $w_i$ and $b_i$ denote the weight and bias of $i$-th layer of micro-MLP, $\|$ indicates the concatenation operation, $L$ determines the depth of micro-MLP, $|\cdot|$ gets the size of the variables. A visual illustration of this process is shown in the left part of Fig.\ref{fig:framework}. 

$P_{feed}$ is then fed into the micro-MLP and the feature interaction is realized via the concatenation of the output of each layer:

\begin{align}
    & h_{0} = P_{feed} \\
    & h_{i} = \sigma(w_{i-1} \otimes h_{i-1} + b_{i-1}), \quad i = 1,2, \cdots , L  \\
    & F(u_{o'}, m_{o}) = H(P_{induction}, P_{feed}) = {\Big\Arrowvert_{i=1}^{L}} h_{i}
\end{align}
where $\otimes$ denotes the matrix multiplication, $\sigma$ indicates the activation function, $H$ denotes the co-action unit with vector inputs $P_{induction}$ and $P_{feed}$ instead of original symbol $F$ whose inputs are features $u_{o'}$ and $m_{o}$.

For sequential features like user behavior history $P_{seq}=\{P_{b(t)}\}_{t=1}^{T}$, the co-action unit is applied to each click behavior followed by a sum-pooling over the sequences:

\begin{equation}
H(P_{induction}, P_{seq}) =  H(P_{induction},\sum_{t=1}^{T}P_{b(t)})
\end{equation}

In our implementation, $P_{induction}$ gets the information from item features while $P_{feed}$ is from user features. However, $P_{feed}$ can also serve as the parameter of micro-MLP and vice versa for $P_{induction}$. Empirically, in the advertisement system, the candidate items are a small part of all items so that their number is less than that of the items in user click history. Hence we choose $P_{induction}$ as the micro-MLP parameter to reduce the total parameter, which makes the learning procedure easier and more stable.

Note that the number of micro-MLP layers depends on the difficulty of learning. Empirically, a larger feature size usually requires deeper MLP. In fact, FM\cite{rendle2010factorization,guo2017deepfm} can also be regarded as a special case of CAN when the micro-MLP is one layer $1 \times D$ matrix without bias and activation function. 

The proposed co-action unit has at least three advantages compared with other methods. First, different from previous works that use the same latent vectors in different types of inter-field interactions, the co-action unit utilizes the computational ability of micro-MLP and couples two component features $P_{induction}$ and $P_{feed}$ dynamically rather than a fixed model, which provides more capacity to guarantee the disentangled update of two field features. Second, a smaller scale of the parameters needs to learn. 
For instance, considering two features with both $N$ IDs, the parameter scale of their cartesian product should be $O(N^2 \times D)$, where $D$ is the dimension of embeddings. However, by using co-action unit, this scale will decrease to $O(N \times (D' + D))$, where $D'$ is the dimension of the $P_{induction}$ in co-action unit. Fewer parameters are not only conducive to learning but also can effectively reduce the burden of the online system. Third, the co-action unit has a better generalization to new feature combinations compared with cartesian product. Given a new feature combination, the co-action unit still works as long as their embeddings of two sides are trained before. 

\begin{figure}[t]
	\centering
	\includegraphics[width=0.8\linewidth]{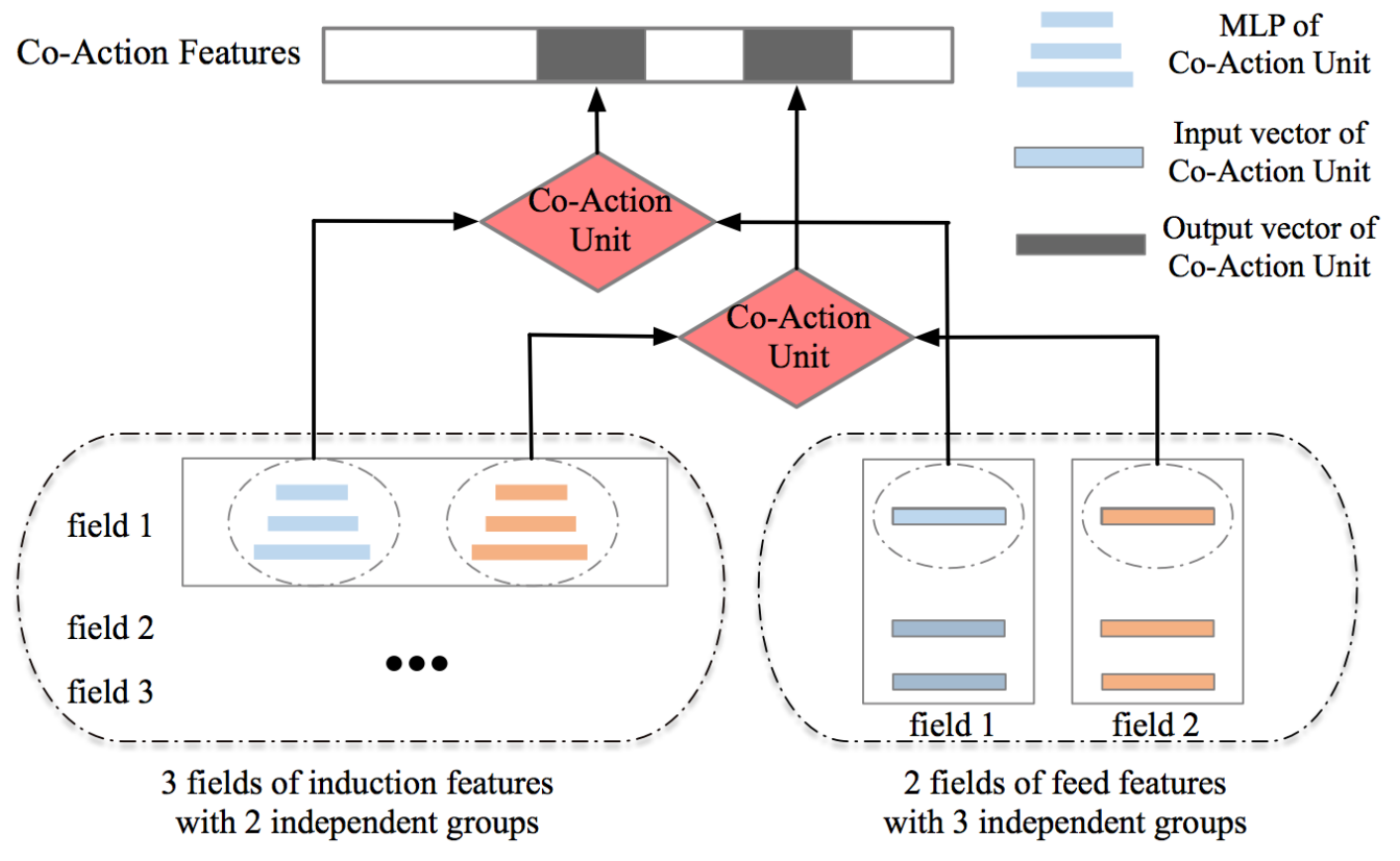}
	\caption{Illustration of combination independence.}
	\label{fig:multi-level}
\end{figure}

\subsection{Multi-order Enhancement} 
\label{sec:multi-order}

The aforementioned feature is formed upon the first-order features. However, feature interaction can be estimated over high-orders. Although the co-action unit can implicitly learn the high-order feature interaction considering the non-linearity of micro-MLP, the learning process is supposed to be difficult because of the sparsity of feature interaction. To this end, we explicitly introduce multi-order information to obtain a polynomial input. This is achieved by applying micro-MLP to different orders of $P_{feed}$:
\begin{align}
	\label{eq:orders}
	H_{\text{Multi-order}}(P_{induction}, P_{feed}) = \sum_{c=1}^C H \big( P_{induction}, (P_{feed})^c \big)
\end{align}
where $C$ is the number of orders. We utilize Tanh to avoid the numerical problem caused by high-order terms. The multi-order enhancement effectively promotes the model's non-linear fitting ability without bringing additional computational and storage costs.

\subsection{Multi-level Independence}
Learning independence is one of the main concerns for feature interaction modeling. To ensure learning independence, we propose a three-level strategy from different aspects.

The first level, parameter independence, which is necessary. As mentioned in Sec.\ref{subsec:co-action_unit}, our approach disentangles the update of representation learning and feature interaction modeling. The parameter independence is the basis of our CAN.

The second level, combination independence, which is recommended. The feature interaction grows linearly as the number of feature combinations increases. Empirically, the target item features like ``item\_id'' and ``category\_id'' are selected as the induction side while the user features are for the feed side. Since one induction side micro-MLP can be combined with several feed sides and vice versa, our approach can easily enlarge the expression capability of the model exponentially. We illustrated this idea in Fig.\ref{fig:multi-level}. Formally, if the induction and feed sides have $Q$ and $S$ groups respectively, the combinations of feature interaction should satisfy:

\begin{align}
\label{eq:emb_dim2}
    &|P_{induction}| = \sum\limits_{s=1}^{S}\sum\limits_{i=0}^{L_s - 1} \big(|w_{i}(s)|+|b_{i}(s)|\big)  \\
    &|P_{feed}| = \sum\limits_{q=1}^{Q}|x(q)|
\end{align}
where $|x(q)|$ is the input dimension of the $q$-th micro-MLP. In the forward pass, the feed feature is divided into several parts to fulfill each micro-MLP.

The third level, orders independence, which is optional. To further improve the flexibility of feature interaction modeling in multi-order inputs, our approach makes different induction side embeddings for different orders. However, the dimension of these embeddings correspondingly increases $C$ times similar to Eq.\ref{eq:emb_dim2}.

The multi-level independence helps the feature interaction modeling but at the same time, brings additional memory access and computations. There is a trade-off between the independence level and deployment cost. Empirically, the higher independence level the model uses, the more training data it needs. In our real system, three levels of independence are used yet only parameter independence is used in public datasets due to the lack of training samples. 


\section{Experiments}

In this section, we present the empirical studies in detail. In Sec.\ref{sec:setting}, we first introduce general experimental settings. The results and discussions are elaborated in Sec.\ref{sec:res}. Sec.\ref{sec:ablation} evaluates the effect of each component via the ablation studies. 

\subsection{Experimental Settings}\label{sec:setting}
\textbf{Dataset.} We conduct experiments on three publicly accessible datasets for CTR prediction task: Amazon, Taobao and Avazu, whose statistics are summarized in Tab.\ref{tab:dataset}.
\begin{itemize}[leftmargin=*]
	\item \textbf{Amazon dataset\footnote{http://jmcauley.ucsd.edu/data/amazon/}} contains product reviews and metadata from Amazon. Among 24 categories of products, we select the Books subset in our experiment. Following previous works \cite{zhou2018din,zhou2019dien,Pi2019MIMN}, we randomly select products not rated by a specific user as negative samples and create corresponding user behavior sequence.
	\item \textbf{Taobao dataset\footnote{https://tianchi.aliyun.com/dataset/dataDetail?dataId=649}} is a set of user behaviors from Taobao’s recommender system. The dataset contains about 1 million users whose behaviors include click, purchase, adding items to the shopping cart, etc. The click behaviors for each user are taken and sorted according to the timestamp to construct the behavior sequence. 
	\item \textbf{Avazu dataset\footnote{https://www.kaggle.com/c/avazu-ctr-prediction}} is a mobile ad dataset including 11 days (10 days for training and 1 day for test) real industrial data provided by Avazu. Different from Amazon and Taobao datasets, we model the feature interaction using discrete features as this dataset contains various data fields, which is suitable to verify the effect of (non-)sequence to feature interaction modeling. During training, data of the 10th day is regarded as the validation set. 
\end{itemize}

\begin{table}[t]
	\caption{Statistics of datasets used in this paper.}
	\label{tab:dataset}
	\begin{tabular}{lccc}
	\toprule[1.5pt]
	Dataset       & Training  & Validation  & Feature Size  \\ 
	\midrule
	Amazon (book) &1086120     &121216        & 912642        \\ 
	Taobao        &691456     &296192       & 5159463            \\ 
	Avazu         &36387240   &403793       & 6763060  \\ 
	\bottomrule[1.5pt]
	\end{tabular}
\end{table}

\textbf{Competitors.} 

To verify the effectiveness of our approach, we compare CAN with state-of-the-art CTR prediction models focusing on feature interaction modeling. 
\begin{itemize}[leftmargin=*]
  	\item \textbf{Cartesian product} is the multiplication of two sets to form the set of all ordered pairs. The former of the produced pair belongs to the first set and the latter is from the second set.
	\item \textbf{DeepFM}~\cite{guo2017deepfm} based on DNN and adopts a product layer to combine the power of factorization machines for recommendation.
	\item \textbf{xDeepFM}~\cite{lian2018xdeepfm} aims to generate feature interactions in an explicit fashion and at the vector-wise level using the proposed Compressed Interaction Network (CIN).
	\item \textbf{FFM \& DeepFFM}~\cite{juan2016field} is a variant of Factorization Machines (FMs) with field aware that can classify large sparse data. DeepFFM appends a DNN term to incorporate high order combination information implicitly.  
	\item \textbf{PNN}~\cite{qu2016product} uses a product layer followed by a fully connected layer to explore high-order feature interactions.
	\item \textbf{NCF}~\cite{he2017neural} presents a neural network architecture to model collaborative filtering between the latent vectors.
	\item \textbf{ONN}~\cite{yang2020operation} proposes Operation-aware Neural Networks which learns different representations for different operations.
	\item \textbf{DIEN} ~\cite{zhou2019dien} designs an interest extractor layer to capture user interests from user behavior sequence. An interest evolving layer is further used to model the interest evolving process.
\end{itemize}

For a fair comparison, DNN is used as the base model (\textbf{CAN-DNN}) so that the differences of these methods (except DIEN) lie on the feature interaction modeling. Meanwhile, we present additional experiments based on DIEN (\textbf{CAN-DIEN}), which is a state-of-the-art method focusing on user interest, to evaluate the promotion of CAN on sequence-oriented modeling.

\textbf{Implementation details.} We implement CAN using Tensorflow. Specifically, for $P_{induction}$, a two-layer MLP is used with input/output dimension set to 16/8 and 8/4. The order of $P_{feed}$ is set to 3. The model parameters are initialized with a Gaussian distribution (with a mean of 0 and standard deviation of 0.01). We use Adam to optimize the training procedure with the batch size set to 128 and the learning rate set to 0.001. A three-layer MLP with layers $200 \times 80 \times 2$ is used for final CTR prediction. The commonly used metric AUC is taken to evaluate the model performance. Note that all experiments are conducted 5 times independently from random train and validation partitions.

\begin{table*}[t]
    \caption{The AUC performance of Cartesian Product and CAN based on DNN in Amazon (book), Taobao and Avazu datasets.}
    \vspace{-2mm}
    \label{tab:cp_exp}
    \resizebox{0.95\textwidth}{!}{
    \begin{tabular}{lccccccc}
    \toprule[1.5pt]
    \multirow{2}{*}{\diagbox{Model}{Dataset}} & \multicolumn{2}{c}{\bf Amazon (book)} & \multicolumn{2}{c}{\bf Taobao} & \multicolumn{2}{c}{\bf Avazu} \\ 
    \cmidrule(lr){2-3} \cmidrule(lr){4-5} \cmidrule(lr){6-7} 
    & AUC (mean ± std) & Parameter & AUC (mean ± std)  & Parameter & AUC (mean ± std)  & Parameter  \\ 
    \midrule
    DNN                  &  0.7640 $\pm$ 0.0007  & 1.0x  &  0.8470 $\pm$ 0.0011  & 1.0x  &  0.7624 $\pm$ 0.0008   & 1.0x  \\
    \midrule
    + Cartesian          &  0.7891 $\pm$ 0.0007 (3.29\% $\uparrow$)  & 17.0x  &  0.8863 $\pm$ 0.0012 (4.64\% $\uparrow$) & 16.5x  &  0.8041 $\pm$ 0.0014 (5.47\% $\uparrow$)  & 21.0x  \\ 
    + CAN                &  0.7935 $\pm$ 0.0007 (3.86\% $\uparrow$) & 3.3x  &  0.8828 $\pm$ 0.0016 (4.23\% $\uparrow$) & 2.6x  &  0.8037 $\pm$ 0.0013 (5.42\% $\uparrow$)  & 3.3x  \\ 
    + CAN + Cartesian    &  0.8054 $\pm$ 0.0007 (5.42\% $\uparrow$) & 18.8x  &  0.8967 $\pm$ 0.0017 (5.87\% $\uparrow$) & 15.3x  &  0.8120 $\pm$ 0.0014 (6.51\% $\uparrow$)  & 23.4x  \\ 
    \bottomrule[1.5pt]
    \end{tabular}}
\end{table*}

\begin{table}[t]
    \caption{The AUC performance on Amazon (book) and Taobao datasets (sequential).}
    \vspace{-2mm}
    \label{tab:result}
    \begin{tabular}{lccc}
    \toprule[1.5pt]
    \multirow{2}{*}{\diagbox{Model}{Dataset}} & Amazon (book) & Taobao  \\ 
    \cmidrule(lr){2-2} \cmidrule(lr){3-3} 
    & AUC (mean ± std) & AUC (mean ± std)  \\ 
    \midrule
    FFM                           &  0.7523 $\pm$ 0.0004     &  0.7918 $\pm$ 0.0016     \\ 
    DNN                           &  0.7640 $\pm$ 0.0007     &  0.8470 $\pm$ 0.0011     \\ 
    DeepFM                        &  0.7682 $\pm$ 0.0005     &  0.8500 $\pm$ 0.0012     \\ 
    DeepFFM                       &  0.7711 $\pm$ 0.0004     &  0.8545 $\pm$ 0.0016     \\ 
    xDeepFM                       &  0.7697 $\pm$ 0.0005     &  0.8573 $\pm$ 0.0012     \\ 
    PNN                           &  0.7801 $\pm$ 0.0002     &  0.8649 $\pm$ 0.0014     \\ 
    NCF                           &  0.7820 $\pm$ 0.0005     &  0.8717 $\pm$ 0.0023     \\ 
    ONN                           &  0.7851 $\pm$ 0.0007     &  0.8752 $\pm$ 0.0011     \\ 
    DIEN                          &  0.8346 $\pm$ 0.0007     &  0.9262 $\pm$ 0.0011     \\ 
    \midrule
    \textbf{CAN-DNN}              & \bf0.7935 $\pm$ 0.0007      & \bf0.8828 $\pm$ 0.0016      \\ 
    \textbf{CAN-DIEN}             & \bf 0.8468 $\pm$ 0.0008  & \bf 0.9388 $\pm$ 0.0013  \\ 
    \bottomrule[1.5pt]
    \end{tabular}
\end{table}

\begin{table}[t]
    \caption{Results on Avazu dataset (non-sequential).}
    \vspace{-2mm}
    \label{tab:general1}
    \begin{tabular}{lc}
    \toprule[1.5pt]
    Model             &  AUC (mean ± std) \\ 
    \midrule
    FFM                      &  0.7580 $\pm$ 0.0014      \\ 
    DNN                      &  0.7624 $\pm$ 0.0008      \\ 
    DeepFM                   &  0.7712 $\pm$ 0.0015      \\ 
    DeepFFM                  &  0.7746 $\pm$ 0.0013      \\ 
    xDeepFM                  &  0.7664 $\pm$ 0.0014      \\ 
    PNN                      &  0.7871 $\pm$ 0.0011      \\ 
    NCF                      &  0.7865 $\pm$ 0.0012      \\ 
    ONN                      &  0.7902 $\pm$ 0.0014      \\ 
    \midrule
    \bf CAN-DNN              &  \bf 0.8037 $\pm$ 0.0013  \\ 
    \bottomrule[1.5pt]
    \end{tabular}
\end{table}

\subsection{Results}\label{sec:res}
\textbf{Overall Performance. }
We report the performance of our proposed CAN and baselines on both three datasets in Tab.\ref{tab:result} and Tab.\ref{tab:general1}. 

Tab.\ref{tab:result} shows the experimental results on Amazon(book) and Taobao dataset. As mentioned in Tab.\ref{tab:result}, CAN-DNN outperforms other state-of-the-art approaches on both datasets and improves AUC by 3.86\% and 4.23\%, respectively, compared with the base model DNN. 
Meanwhile, CAN-DNN surpasses other feature interaction approaches with a large margin, outperforming the strongest feature interaction (non-sequatial) baseline ONN in both two datasets.
Therefore, it validates the effectiveness of our approach on interaction modeling. 
Since the two datasets contain rich user sequential data, sequence-oriented approaches (which are more suitable for real industrial systems) like DIEN performs better than DNN. Thus, we also use DIEN as the base model to evaluate the effects of CAN (CAN-DIEN). The results show that DIEN can still benefit from CAN with 1.46\% and 1.36\% improvement on AUC, respectively. 

Tab.\ref{tab:general1} shows the experimental results on Avazu dataset. Although CAN is designed mainly for real industrial data that contain a lot of behavior sequences, it's still capable of non-sequential input. The Avazu dataset contains 24 data fields among which we select 9 fields to construct 16 kinds of feature combinations. Empirical results show that CAN outperforms all the other approaches.

\textbf{Compare with Cartesian Product. }
Firstly, it's worth noting that, as a pure representation learning method, cartesian product method could achieve better performance compared with other embedding combination methods. 
It indicates that although those combination methods could extract some information of feature interaction, there is still a large gap to direct combination encoding, i.e., cartesian product. 
As shown in Tab.\ref{tab:cp_exp}, CAN achieves comparable results with the cartesian product by only 1/6 parameters. 

Meanwhile, we found there has a strong overlap between CAN and cartesian product in feature interaction, for example, CAN ($+$3.86$\%$) and cartesian product ($+$3.28$\%$) only brings 5.42\% lift on Amazon(book) dataset. The experimental results on Taobao dataset and Avazu dataset show a similar result and this reveals that CAN can effectively model the feature co-occurrence.

\begin{figure}[t]
	\centering
	\includegraphics[width=0.9\linewidth]{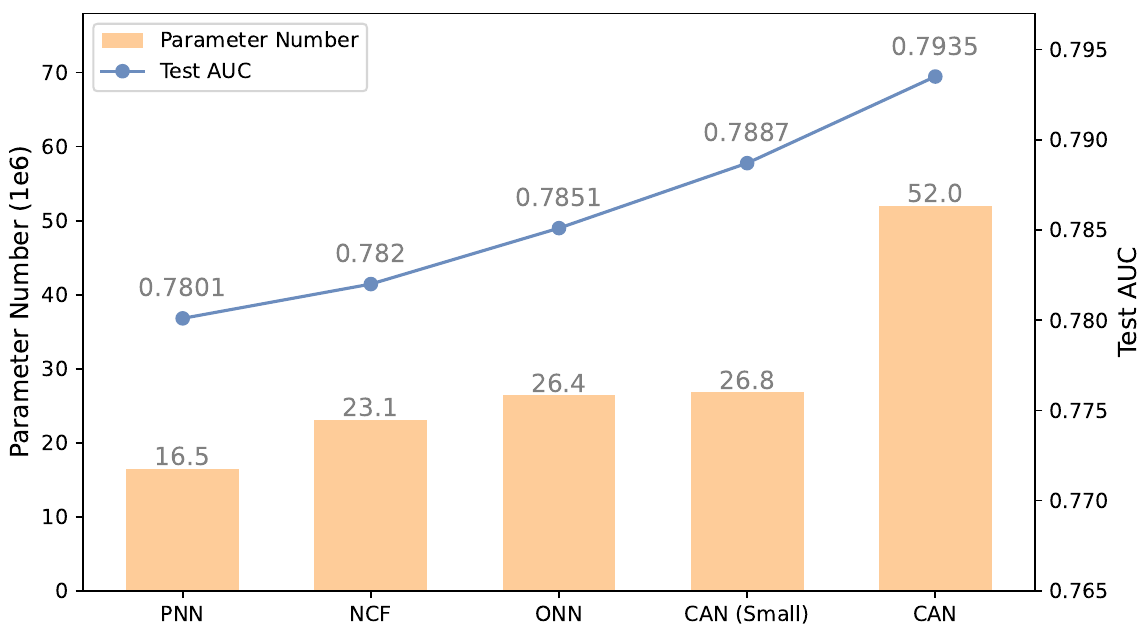}
	\caption{Parameter number (orange bar) and corresponding test AUC (blue line) of different methods.}
	\label{fig:parameter}
\end{figure}

\textbf{Parameter Number Analysis. }
Fig.\ref{fig:parameter} shows the parameter number and corresponding test AUC of different methods on Amazon book dataset. In CAN (Small) model, we set the dimensions of tensor in co-action unit and extra embedding layers to a small value to keep similar parameter number with other methods. As shown in Fig.\ref{fig:parameter}, CAN (Small) significantly outperforms other baselines which have similar number of parameters, such as ONN. It proves that the improvement of our CAN does not come from the amplified parameter number but from feature interaction modeling by co-action unit. Moreover, the performance can be further improved by increasing the parameter of CAN since the feature interaction modeling is closely related to the learning of micro-MLP.

\subsection{Ablation Study}\label{sec:ablation}
To investigate the effect of each component, we conduct several ablation studies, which are shown in Tab.\ref{tab:ablation}.

\textbf{MLP Layer. (mentioned in Sec.~\ref{subsec:co-action_unit})} First, we show the influences of $MLP_{can}$ architecture to the feature interaction modeling. Specifically, we train models with a different number of MLP layers: 1 layer, 2 and 3 layers. The input/output dimension of each layer is set to 16/8, 8/4 and 4/4, respectively. In general, deeper MLP leads to higher performance. However, when the number of layers grows to 3, there is a decrease in AUC. The main reason may be that the network is not well trained as more complex network architecture usually requires more training data to get convergence. 

\textbf{Activation Function.} Second, we study the influences of the activation function. As can be seen from Tab.\ref{tab:ablation}, the non-linearity improves AUC by 0.23\%. The Tanh activation function plays the role of a normalizer to avoid numerical issues in high orders and helps the model to train stably. 

\textbf{Multi-order Enhancement (mentioned in Sec.~\ref{sec:multi-order}).} Third, we evaluate the influences of multi orders. Based on 1st order term, 2nd, 3rd and 4th order terms are added, gradually. From 1st order to 3rd order, AUC promotes a lot. Afterward, as the order grows, the gap starts to shrink even causing a negative effect. The multi orders have a marginal effect on the performance gain so that 2 or 3 power terms are proper in 
reality.

\begin{table}[t]
	\caption{Ablation studies on Amazon(book) dataset.}
	\label{tab:ablation}
	\begin{tabular}{lc}
	\toprule[1.5pt]
	Components          & AUC (mean ± std)  \\ 
	\midrule
	MLP in Co-Action Unit \quad \quad \\
	\quad \quad 1 layer            &0.7889 $\pm$ 0.0007               \\ 
	\quad \quad 2 layers           &0.7935 $\pm$ 0.0007               \\ 
	\quad \quad 3 layers           &0.7913 $\pm$ 0.0013               \\ 
	\midrule
	Activation Function \\
	\quad \quad w/o activation          &0.7917 $\pm$ 0.0008               \\ 
	\quad \quad w/\phantom{o} Tanh        &0.7935 $\pm$ 0.0007               \\ 
	\midrule
	Multi-order Enhancement \\
	\quad \quad order=1          &0.7902 $\pm$ 0.0008               \\ 
	\quad \quad order=2          &0.7921 $\pm$ 0.0012               \\ 
	\quad \quad order=3  		 &0.7935 $\pm$ 0.0007               \\ 
	\quad \quad order=4          &0.7934 $\pm$ 0.0014               \\ 
	\bottomrule[1.5pt]
	\end{tabular}
\end{table}

\section{Industrial Experience}\label{sec:industrial}
In this section, we share the industrial experiences of feature interaction modeling in our display advertising system.

Cartesian product is the most direct way in feature interaction modeling as discussed in previous sections. Nevertheless, cartesian product usually leads to heavy resource consumption. On one hand, the model size will expand at an extremely fast rate. The oversized model brings great challenges for storage and network transmission, which further affects the real-time update of the model. On the other hand, it increases the embedding look-up operations in application request as the features are increased at the input stage and this results in latency for system response. 

Existing approaches are more friendly in industrial deployment. However, we also noticed that under the scale of the billions of data, the promotions were very limited compared with cartesian product. Meanwhile, simply increasing the parameter space like expanding the embedding size did not bring additional improvements.

CAN is designed as a new feature interaction modeling scheme under this situation. In our advertising system, 21 kinds of features including 6 ad features (e.g., ad\_id, item\_id, shop\_id, etc.) and 15 kinds of user features (e.g., item\_history, shop\_history, etc.) are selected to construct feature interaction. We notice that CAN can achieve comparable performance compared with cartesian product with only one-tenth of model size.

As mentioned in Sec.\ref{sec:method}, given the 21 kinds of features, CAN allocates additional 21 embeddings due to feature interaction independence. As the user features are mostly behavior sequences with a length of more than 100, additional memory access is required which causes much response latency. Moreover, the computational costs of feature interaction grow linearly according to the number of feature combinations which also brings considerable response latency to our system. To exert all energy of CAN, many efforts are devoted to reducing the response latency. In the industrial deployment, we optimize CAN from three aspects:

\begin{itemize}[leftmargin=*]
	\item \textbf{Sequence cut-off.} The length of 16 kinds user features ranges from 50 to 200. We elegantly apply sequence cut-off to reduce memory cost, e.g., all user behavior sequences with length 200 are truncated at length 50. The most recent behaviors are kept. The sequence cut-off promotes the QPS (Query Per Second) by 20\% yet results in a 0.1\% decrease of AUC, which is acceptable.
	\item \textbf{Combination reduction.} 6 ad features and 15 user features can obtain up to 90 combinations which is a heavy burden. Empirically, the combinations of the same kind of ad and user feature can better model the co-occurrence. According to this principle, we keep the combinations like ``item\_id'', ``item\_click\_history'', ``category\_id'', ``category\_click\_history'' and remove some irrelevant combinations. In this way, the number of feature combinations reduces from 90 to 48 which brings 30\% QPS improvement.
	\item \textbf{Computational kernel optimization.} The feature interaction computation refers to a time-consuming large matrix multiplication between $P_{induction}$ and $P_{feed}$ with shape of [batch\_size $\times$ $M$ $\times$  dim\_in $\times$ dim\_out] $\times$ [batch\_size $\times$ $M$ $\times$  $T \times$ dim\_in], where $M$, $T$, dim\_in and dim\_out denote the number of feature interaction, length of user behavior sequence, MLP input and output dimension, respectively. In our case, the dim\_in and dim\_out are not commonly used shape so that such matrix multiplication is not well optimized by the BLAS (Basic Linear Algebra Subprograms). To solve this, the internal calculation logic is rewritten which brings 60\% QPS lift. Besides, we make a kernel fusion which combines several operations (such as Matmul and Tanh) as one to reduce the GPU memory I/O consumption. By doing so, the intermediate GPU memory writing of the matrix multiplication output is avoided, which brings another 47\% QPS lift.  
\end{itemize}

The series of optimizations make CAN capable of online serving stably in the main traffic of our advertisement system. In the case of our practice, the CTR prediction step of CAN takes about 10 ms and the system can handle nearly 1.3K QPS per Tesla T4 GPU.

\begin{table}[t]
    \caption{The performance in real online advertising system.}
	\label{tab:indus}
	\begin{tabular}{lcc}
	\toprule[1.5pt]
	Scene         & CTR & RPM \\
	\midrule
	Homepage Advertising      & +11.4\%  & +8.8\% \\
	Post-purchase Advertising & +12.5\%  & +7.5\% \\
	\bottomrule[1.5pt]
	\end{tabular}
\end{table}

\subsection{Offline and Online Results} 
In all public datasets, CAN has an absolute increase of about 1.2\% in AUC compared with DIEN as shown in Tab.\ref{tab:result}. Meanwhile, CAN also brings an additional 1.2\% GAUC (Group AUC) lift in our 30-days industrial data compared with our online serving model. Tab.\ref{tab:indus} shows the online A/B test results of CAN on our two main scenes in feeds recommendation, Homepage Advertising and Post-purchase Advertising between Jul. 2020 and Oct. 2020. CAN achieves 11.4\%/12.5\% CTR and 8.8\%/7.5\% RPM promotions respectively, which are considerable in industrial practice.


\section{Conclusion}

In this paper, we stress the importance of feature interaction modeling, which is not fully explored by previous works. Inspired by cartesian product, we propose a new feature interaction paradigm using a specially designed network, Co-Action Network (CAN). The CAN disentangles the representation learning and feature interaction modeling via a flexible module, co-action unit. Moreover, multi-order enhancement and multi-level independence are introduced in co-action unit to further promote the ability of feature interaction modeling. The experiments show that the CAN outperforms the previous works. CAN has been deployed in the display advertisement system in Alibaba and served the main traffic. We believe this work has pushed feature interaction learning a step forward and multi-feature and lightweight interaction modeling will be further explored in the future.

\bibliographystyle{ACM-Reference-Format}
\balance
\bibliography{main}


\begin{thebibliography}{28}


\ifx \showCODEN    \undefined \def \showCODEN     #1{\unskip}     \fi
\ifx \showDOI      \undefined \def \showDOI       #1{#1}\fi
\ifx \showISBNx    \undefined \def \showISBNx     #1{\unskip}     \fi
\ifx \showISBNxiii \undefined \def \showISBNxiii  #1{\unskip}     \fi
\ifx \showISSN     \undefined \def \showISSN      #1{\unskip}     \fi
\ifx \showLCCN     \undefined \def \showLCCN      #1{\unskip}     \fi
\ifx \shownote     \undefined \def \shownote      #1{#1}          \fi
\ifx \showarticletitle \undefined \def \showarticletitle #1{#1}   \fi
\ifx \showURL      \undefined \def \showURL       {\relax}        \fi
\providecommand\bibfield[2]{#2}
\providecommand\bibinfo[2]{#2}
\providecommand\natexlab[1]{#1}
\providecommand\showeprint[2][]{arXiv:#2}

\bibitem[\protect\citeauthoryear{Bruna, Zaremba, Szlam, and LeCun}{Bruna
  et~al\mbox{.}}{2014}]%
        {BrunaZSL13SpectralGraph}
\bibfield{author}{\bibinfo{person}{Joan Bruna}, \bibinfo{person}{Wojciech
  Zaremba}, \bibinfo{person}{Arthur Szlam}, {and} \bibinfo{person}{Yann
  LeCun}.} \bibinfo{year}{2014}\natexlab{}.
\newblock \showarticletitle{Spectral Networks and Locally Connected Networks on
  Graphs}. In \bibinfo{booktitle}{\emph{ICLR}}.
\newblock


\bibitem[\protect\citeauthoryear{Chan, Zhang, Chen, Gao, Zhang, Zhao, and
  Yan}{Chan et~al\mbox{.}}{2020}]%
        {chan2020selection}
\bibfield{author}{\bibinfo{person}{Zhangming Chan}, \bibinfo{person}{Yuchi
  Zhang}, \bibinfo{person}{Xiuying Chen}, \bibinfo{person}{Shen Gao},
  \bibinfo{person}{Zhiqiang Zhang}, \bibinfo{person}{Dongyan Zhao}, {and}
  \bibinfo{person}{Rui Yan}.} \bibinfo{year}{2020}\natexlab{}.
\newblock \showarticletitle{Selection and Generation: Learning towards
  Multi-Product Advertisement Post Generation}. In
  \bibinfo{booktitle}{\emph{EMNLP}}. \bibinfo{pages}{3818--3829}.
\newblock


\bibitem[\protect\citeauthoryear{Cheng, Koc, Harmsen, Shaked, Chandra, Aradhye,
  Anderson, Corrado, Chai, Ispir, et~al\mbox{.}}{Cheng et~al\mbox{.}}{2016}]%
        {cheng2016wide}
\bibfield{author}{\bibinfo{person}{Heng-Tze Cheng}, \bibinfo{person}{Levent
  Koc}, \bibinfo{person}{Jeremiah Harmsen}, \bibinfo{person}{Tal Shaked},
  \bibinfo{person}{Tushar Chandra}, \bibinfo{person}{Hrishi Aradhye},
  \bibinfo{person}{Glen Anderson}, \bibinfo{person}{Greg Corrado},
  \bibinfo{person}{Wei Chai}, \bibinfo{person}{Mustafa Ispir}, {et~al\mbox{.}}}
  \bibinfo{year}{2016}\natexlab{}.
\newblock \showarticletitle{Wide \& deep learning for recommender systems}. In
  \bibinfo{booktitle}{\emph{DLRS}}. \bibinfo{pages}{7--10}.
\newblock


\bibitem[\protect\citeauthoryear{Feng, Lv, Shen, Wang, Sun, Zhu, and Yang}{Feng
  et~al\mbox{.}}{2019}]%
        {FengLSWSZY19DSIN}
\bibfield{author}{\bibinfo{person}{Yufei Feng}, \bibinfo{person}{Fuyu Lv},
  \bibinfo{person}{Weichen Shen}, \bibinfo{person}{Menghan Wang},
  \bibinfo{person}{Fei Sun}, \bibinfo{person}{Yu Zhu}, {and}
  \bibinfo{person}{Keping Yang}.} \bibinfo{year}{2019}\natexlab{}.
\newblock \showarticletitle{Deep Session Interest Network for Click-Through
  Rate Prediction}. In \bibinfo{booktitle}{\emph{IJCAI}}.
  \bibinfo{pages}{2301--2307}.
\newblock


\bibitem[\protect\citeauthoryear{Gori, Monfardini, and Scarselli}{Gori
  et~al\mbox{.}}{2005}]%
        {Gori2005GNN}
\bibfield{author}{\bibinfo{person}{Marco Gori}, \bibinfo{person}{Gabriele
  Monfardini}, {and} \bibinfo{person}{Franco Scarselli}.}
  \bibinfo{year}{2005}\natexlab{}.
\newblock \showarticletitle{A new model for learning in graph domains}. In
  \bibinfo{booktitle}{\emph{IJCNN}}, Vol.~\bibinfo{volume}{2}.
  \bibinfo{pages}{729--734}.
\newblock


\bibitem[\protect\citeauthoryear{Guo, Tang, Ye, Li, and He}{Guo
  et~al\mbox{.}}{2017}]%
        {guo2017deepfm}
\bibfield{author}{\bibinfo{person}{Huifeng Guo}, \bibinfo{person}{Ruiming
  Tang}, \bibinfo{person}{Yunming Ye}, \bibinfo{person}{Zhenguo Li}, {and}
  \bibinfo{person}{Xiuqiang He}.} \bibinfo{year}{2017}\natexlab{}.
\newblock \showarticletitle{Deepfm: a factorization-machine based neural
  network for ctr prediction}. In \bibinfo{booktitle}{\emph{IJCAI}}.
  \bibinfo{pages}{2782--2788}.
\newblock


\bibitem[\protect\citeauthoryear{He, Liao, Zhang, Nie, Hu, and Chua}{He
  et~al\mbox{.}}{2017}]%
        {he2017neural}
\bibfield{author}{\bibinfo{person}{Xiangnan He}, \bibinfo{person}{Lizi Liao},
  \bibinfo{person}{Hanwang Zhang}, \bibinfo{person}{Liqiang Nie},
  \bibinfo{person}{Xia Hu}, {and} \bibinfo{person}{Tat-Seng Chua}.}
  \bibinfo{year}{2017}\natexlab{}.
\newblock \showarticletitle{Neural collaborative filtering}. In
  \bibinfo{booktitle}{\emph{WWW}}. \bibinfo{pages}{173--182}.
\newblock


\bibitem[\protect\citeauthoryear{Hu, Chan, Liu, Zhao, Ma, and Yan}{Hu
  et~al\mbox{.}}{2019}]%
        {hu2019gsn}
\bibfield{author}{\bibinfo{person}{Wenpeng Hu}, \bibinfo{person}{Zhangming
  Chan}, \bibinfo{person}{Bing Liu}, \bibinfo{person}{Dongyan Zhao},
  \bibinfo{person}{Jinwen Ma}, {and} \bibinfo{person}{Rui Yan}.}
  \bibinfo{year}{2019}\natexlab{}.
\newblock \showarticletitle{GSN: A graph-structured network for multi-party
  dialogues}.
\newblock \bibinfo{journal}{\emph{IJCAI}} (\bibinfo{year}{2019}),
  \bibinfo{pages}{5010--5016}.
\newblock


\bibitem[\protect\citeauthoryear{Juan, Zhuang, Chin, and Lin}{Juan
  et~al\mbox{.}}{2016}]%
        {juan2016field}
\bibfield{author}{\bibinfo{person}{Yuchin Juan}, \bibinfo{person}{Yong Zhuang},
  \bibinfo{person}{Wei-Sheng Chin}, {and} \bibinfo{person}{Chih-Jen Lin}.}
  \bibinfo{year}{2016}\natexlab{}.
\newblock \showarticletitle{Field-aware factorization machines for CTR
  prediction}. In \bibinfo{booktitle}{\emph{RecSys}}. \bibinfo{pages}{43--50}.
\newblock


\bibitem[\protect\citeauthoryear{Kipf and Welling}{Kipf and Welling}{2017}]%
        {KipfW17GCN}
\bibfield{author}{\bibinfo{person}{Thomas~N. Kipf} {and} \bibinfo{person}{Max
  Welling}.} \bibinfo{year}{2017}\natexlab{}.
\newblock \showarticletitle{Semi-Supervised Classification with Graph
  Convolutional Networks}. In \bibinfo{booktitle}{\emph{ICLR}}.
\newblock


\bibitem[\protect\citeauthoryear{Li, Liu, Wu, Xu, Zhao, Huang, Kang, Chen, Li,
  and Lee}{Li et~al\mbox{.}}{2019}]%
        {Li2019MIND}
\bibfield{author}{\bibinfo{person}{Chao Li}, \bibinfo{person}{Zhiyuan Liu},
  \bibinfo{person}{Mengmeng Wu}, \bibinfo{person}{Yuchi Xu},
  \bibinfo{person}{Huan Zhao}, \bibinfo{person}{Pipei Huang},
  \bibinfo{person}{Guoliang Kang}, \bibinfo{person}{Qiwei Chen},
  \bibinfo{person}{Wei Li}, {and} \bibinfo{person}{Dik~Lun Lee}.}
  \bibinfo{year}{2019}\natexlab{}.
\newblock \showarticletitle{Multi-Interest Network with Dynamic Routing for
  Recommendation at Tmall}. In \bibinfo{booktitle}{\emph{CIKM}}.
  \bibinfo{pages}{2615--2623}.
\newblock


\bibitem[\protect\citeauthoryear{Lian, Zhou, Zhang, Chen, Xie, and Sun}{Lian
  et~al\mbox{.}}{2018}]%
        {lian2018xdeepfm}
\bibfield{author}{\bibinfo{person}{Jianxun Lian}, \bibinfo{person}{Xiaohuan
  Zhou}, \bibinfo{person}{Fuzheng Zhang}, \bibinfo{person}{Zhongxia Chen},
  \bibinfo{person}{Xing Xie}, {and} \bibinfo{person}{Guangzhong Sun}.}
  \bibinfo{year}{2018}\natexlab{}.
\newblock \showarticletitle{xdeepfm: Combining explicit and implicit feature
  interactions for recommender systems}. In \bibinfo{booktitle}{\emph{SIGKDD}}.
  \bibinfo{pages}{1754--1763}.
\newblock


\bibitem[\protect\citeauthoryear{Pi, Bian, Zhou, Zhu, and Gai}{Pi
  et~al\mbox{.}}{2019}]%
        {Pi2019MIMN}
\bibfield{author}{\bibinfo{person}{Qi Pi}, \bibinfo{person}{Weijie Bian},
  \bibinfo{person}{Guorui Zhou}, \bibinfo{person}{Xiaoqiang Zhu}, {and}
  \bibinfo{person}{Kun Gai}.} \bibinfo{year}{2019}\natexlab{}.
\newblock \showarticletitle{Practice on Long Sequential User Behavior Modeling
  for Click-through Rate Prediction}. In \bibinfo{booktitle}{\emph{SIGKDD}}.
  \bibinfo{pages}{1059--1068}.
\newblock


\bibitem[\protect\citeauthoryear{Qu, Cai, Ren, Zhang, Yu, Wen, and Wang}{Qu
  et~al\mbox{.}}{2016}]%
        {qu2016product}
\bibfield{author}{\bibinfo{person}{Yanru Qu}, \bibinfo{person}{Han Cai},
  \bibinfo{person}{Kan Ren}, \bibinfo{person}{Weinan Zhang},
  \bibinfo{person}{Yong Yu}, \bibinfo{person}{Ying Wen}, {and}
  \bibinfo{person}{Jun Wang}.} \bibinfo{year}{2016}\natexlab{}.
\newblock \showarticletitle{Product-based neural networks for user response
  prediction}. In \bibinfo{booktitle}{\emph{ICDM}}.
  \bibinfo{pages}{1149--1154}.
\newblock


\bibitem[\protect\citeauthoryear{Qu, Fang, Zhang, Tang, Niu, Guo, Yu, and
  He}{Qu et~al\mbox{.}}{2018}]%
        {qu2018product}
\bibfield{author}{\bibinfo{person}{Yanru Qu}, \bibinfo{person}{Bohui Fang},
  \bibinfo{person}{Weinan Zhang}, \bibinfo{person}{Ruiming Tang},
  \bibinfo{person}{Minzhe Niu}, \bibinfo{person}{Huifeng Guo},
  \bibinfo{person}{Yong Yu}, {and} \bibinfo{person}{Xiuqiang He}.}
  \bibinfo{year}{2018}\natexlab{}.
\newblock \showarticletitle{Product-based neural networks for user response
  prediction over multi-field categorical data}.
\newblock \bibinfo{journal}{\emph{ACM Transactions on Information Systems
  (TOIS)}} \bibinfo{volume}{37}, \bibinfo{number}{1} (\bibinfo{year}{2018}),
  \bibinfo{pages}{1--35}.
\newblock


\bibitem[\protect\citeauthoryear{Rendle}{Rendle}{2010}]%
        {rendle2010factorization}
\bibfield{author}{\bibinfo{person}{Steffen Rendle}.}
  \bibinfo{year}{2010}\natexlab{}.
\newblock \showarticletitle{Factorization machines}. In
  \bibinfo{booktitle}{\emph{ICDM}}. \bibinfo{pages}{995--1000}.
\newblock


\bibitem[\protect\citeauthoryear{Sheng, Zhao, Zhou, Ding, Dai, Luo, Yang, Lv,
  Zhang, Deng, and Zhu}{Sheng et~al\mbox{.}}{2021}]%
        {DBLP:conf/cikm/ShengZZDDLYLZDZ21}
\bibfield{author}{\bibinfo{person}{Xiang{-}Rong Sheng}, \bibinfo{person}{Liqin
  Zhao}, \bibinfo{person}{Guorui Zhou}, \bibinfo{person}{Xinyao Ding},
  \bibinfo{person}{Binding Dai}, \bibinfo{person}{Qiang Luo},
  \bibinfo{person}{Siran Yang}, \bibinfo{person}{Jingshan Lv},
  \bibinfo{person}{Chi Zhang}, \bibinfo{person}{Hongbo Deng}, {and}
  \bibinfo{person}{Xiaoqiang Zhu}.} \bibinfo{year}{2021}\natexlab{}.
\newblock \showarticletitle{One Model to Serve All: Star Topology Adaptive
  Recommender for Multi-Domain {CTR} Prediction}. In
  \bibinfo{booktitle}{\emph{CIKM}}. \bibinfo{pages}{4104--4113}.
\newblock


\bibitem[\protect\citeauthoryear{Shi, Hu, Zhao, and Yu}{Shi
  et~al\mbox{.}}{2019}]%
        {ShiHZY2019HINEmbedding}
\bibfield{author}{\bibinfo{person}{Chuan Shi}, \bibinfo{person}{Binbin Hu},
  \bibinfo{person}{Wayne~Xin Zhao}, {and} \bibinfo{person}{Philip~S. Yu}.}
  \bibinfo{year}{2019}\natexlab{}.
\newblock \showarticletitle{Heterogeneous Information Network Embedding for
  Recommendation}.
\newblock \bibinfo{journal}{\emph{{IEEE} TKDE}} \bibinfo{volume}{31},
  \bibinfo{number}{2} (\bibinfo{year}{2019}), \bibinfo{pages}{357--370}.
\newblock


\bibitem[\protect\citeauthoryear{Song, Shi, Xiao, Duan, Xu, Zhang, and
  Tang}{Song et~al\mbox{.}}{2019}]%
        {song2019autoint}
\bibfield{author}{\bibinfo{person}{Weiping Song}, \bibinfo{person}{Chence Shi},
  \bibinfo{person}{Zhiping Xiao}, \bibinfo{person}{Zhijian Duan},
  \bibinfo{person}{Yewen Xu}, \bibinfo{person}{Ming Zhang}, {and}
  \bibinfo{person}{Jian Tang}.} \bibinfo{year}{2019}\natexlab{}.
\newblock \showarticletitle{Autoint: Automatic feature interaction learning via
  self-attentive neural networks}. In \bibinfo{booktitle}{\emph{CIKM}}.
  \bibinfo{pages}{1161--1170}.
\newblock


\bibitem[\protect\citeauthoryear{Sun, Han, Yan, Yu, and Wu}{Sun
  et~al\mbox{.}}{2011}]%
        {SunHYYW2011PathSim}
\bibfield{author}{\bibinfo{person}{Yizhou Sun}, \bibinfo{person}{Jiawei Han},
  \bibinfo{person}{Xifeng Yan}, \bibinfo{person}{Philip~S. Yu}, {and}
  \bibinfo{person}{Tianyi Wu}.} \bibinfo{year}{2011}\natexlab{}.
\newblock \showarticletitle{PathSim: Meta Path-Based Top-K Similarity Search in
  Heterogeneous Information Networks}.
\newblock \bibinfo{journal}{\emph{Proc. VLDB Endow.}} \bibinfo{volume}{4},
  \bibinfo{number}{11} (\bibinfo{year}{2011}), \bibinfo{pages}{992--1003}.
\newblock


\bibitem[\protect\citeauthoryear{Vaswani, Shazeer, Parmar, Uszkoreit, Jones,
  Gomez, Kaiser, and Polosukhin}{Vaswani et~al\mbox{.}}{2017}]%
        {VaswaniSPUJGKP2017Transformer}
\bibfield{author}{\bibinfo{person}{Ashish Vaswani}, \bibinfo{person}{Noam
  Shazeer}, \bibinfo{person}{Niki Parmar}, \bibinfo{person}{Jakob Uszkoreit},
  \bibinfo{person}{Llion Jones}, \bibinfo{person}{Aidan~N. Gomez},
  \bibinfo{person}{Lukasz Kaiser}, {and} \bibinfo{person}{Illia Polosukhin}.}
  \bibinfo{year}{2017}\natexlab{}.
\newblock \showarticletitle{Attention is All you Need}. In
  \bibinfo{booktitle}{\emph{NIPS}}. \bibinfo{pages}{5998--6008}.
\newblock


\bibitem[\protect\citeauthoryear{Velickovic, Cucurull, Casanova, Romero,
  Li{\`{o}}, and Bengio}{Velickovic et~al\mbox{.}}{2018}]%
        {VelickovicCCRLB18GAT}
\bibfield{author}{\bibinfo{person}{Petar Velickovic}, \bibinfo{person}{Guillem
  Cucurull}, \bibinfo{person}{Arantxa Casanova}, \bibinfo{person}{Adriana
  Romero}, \bibinfo{person}{Pietro Li{\`{o}}}, {and} \bibinfo{person}{Yoshua
  Bengio}.} \bibinfo{year}{2018}\natexlab{}.
\newblock \showarticletitle{Graph Attention Networks}. In
  \bibinfo{booktitle}{\emph{ICLR}}.
\newblock


\bibitem[\protect\citeauthoryear{Wang, Fu, Fu, and Wang}{Wang
  et~al\mbox{.}}{2017}]%
        {WangFFW2017DCN}
\bibfield{author}{\bibinfo{person}{Ruoxi Wang}, \bibinfo{person}{Bin Fu},
  \bibinfo{person}{Gang Fu}, {and} \bibinfo{person}{Mingliang Wang}.}
  \bibinfo{year}{2017}\natexlab{}.
\newblock \showarticletitle{Deep {\&} Cross Network for Ad Click Predictions}.
  In \bibinfo{booktitle}{\emph{ADKDD'17}}. \bibinfo{pages}{12:1--12:7}.
\newblock


\bibitem[\protect\citeauthoryear{Xiao, Ye, He, Zhang, Wu, and Chua}{Xiao
  et~al\mbox{.}}{2017}]%
        {xiao2017attentional}
\bibfield{author}{\bibinfo{person}{Jun Xiao}, \bibinfo{person}{Hao Ye},
  \bibinfo{person}{Xiangnan He}, \bibinfo{person}{Hanwang Zhang},
  \bibinfo{person}{Fei Wu}, {and} \bibinfo{person}{Tat-Seng Chua}.}
  \bibinfo{year}{2017}\natexlab{}.
\newblock \showarticletitle{Attentional factorization machines: Learning the
  weight of feature interactions via attention networks}.
\newblock \bibinfo{journal}{\emph{arXiv preprint arXiv:1708.04617}}
  (\bibinfo{year}{2017}).
\newblock


\bibitem[\protect\citeauthoryear{Yang, Xu, Shen, Shen, and Zhao}{Yang
  et~al\mbox{.}}{2020}]%
        {yang2020operation}
\bibfield{author}{\bibinfo{person}{Yi Yang}, \bibinfo{person}{Baile Xu},
  \bibinfo{person}{Shaofeng Shen}, \bibinfo{person}{Furao Shen}, {and}
  \bibinfo{person}{Jian Zhao}.} \bibinfo{year}{2020}\natexlab{}.
\newblock \showarticletitle{Operation-aware Neural Networks for user response
  prediction}.
\newblock \bibinfo{journal}{\emph{Neural Networks}}  \bibinfo{volume}{121}
  (\bibinfo{year}{2020}), \bibinfo{pages}{161--168}.
\newblock


\bibitem[\protect\citeauthoryear{Zhao, Yao, Li, Song, and Lee}{Zhao
  et~al\mbox{.}}{2017}]%
        {ZhaoYLSL2017HINFusion}
\bibfield{author}{\bibinfo{person}{Huan Zhao}, \bibinfo{person}{Quanming Yao},
  \bibinfo{person}{Jianda Li}, \bibinfo{person}{Yangqiu Song}, {and}
  \bibinfo{person}{Dik~Lun Lee}.} \bibinfo{year}{2017}\natexlab{}.
\newblock \showarticletitle{Meta-Graph Based Recommendation Fusion over
  Heterogeneous Information Networks}. In \bibinfo{booktitle}{\emph{SIGKDD}}.
  \bibinfo{pages}{635--644}.
\newblock


\bibitem[\protect\citeauthoryear{Zhou, Mou, Fan, Pi, Bian, Zhou, Zhu, and
  Gai}{Zhou et~al\mbox{.}}{2019}]%
        {zhou2019dien}
\bibfield{author}{\bibinfo{person}{Guorui Zhou}, \bibinfo{person}{Na Mou},
  \bibinfo{person}{Ying Fan}, \bibinfo{person}{Qi Pi}, \bibinfo{person}{Weijie
  Bian}, \bibinfo{person}{Chang Zhou}, \bibinfo{person}{Xiaoqiang Zhu}, {and}
  \bibinfo{person}{Kun Gai}.} \bibinfo{year}{2019}\natexlab{}.
\newblock \showarticletitle{Deep Interest Evolution Network for Click-Through
  Rate Prediction}. In \bibinfo{booktitle}{\emph{AAAI}}.
  \bibinfo{pages}{5941--5948}.
\newblock


\bibitem[\protect\citeauthoryear{Zhou, Zhu, Song, Fan, Zhu, Ma, Yan, Jin, Li,
  and Gai}{Zhou et~al\mbox{.}}{2018}]%
        {zhou2018din}
\bibfield{author}{\bibinfo{person}{Guorui Zhou}, \bibinfo{person}{Xiaoqiang
  Zhu}, \bibinfo{person}{Chenru Song}, \bibinfo{person}{Ying Fan},
  \bibinfo{person}{Han Zhu}, \bibinfo{person}{Xiao Ma},
  \bibinfo{person}{Yanghui Yan}, \bibinfo{person}{Junqi Jin},
  \bibinfo{person}{Han Li}, {and} \bibinfo{person}{Kun Gai}.}
  \bibinfo{year}{2018}\natexlab{}.
\newblock \showarticletitle{Deep interest network for click-through rate
  prediction}. In \bibinfo{booktitle}{\emph{SIGKDD}}.
  \bibinfo{pages}{1059--1068}.
\newblock


\end{thebibliography}

\end{document}